\documentclass[a4paper,twoside, 11pt]{article}
\usepackage[dvips]{graphicx}
\usepackage{color}
\def\gtap{\raisebox{-.55ex}{\rlap{$\sim$}} \raisebox{.4ex}{$>$}}
\def\gsim{\mathrel{\gtap}}
\title{Strong coupling in massive gravity by direct calculation}
\author{Aur\`{e}le Aubert\\ 
{\em Institute of Theoretical Physics,
Swiss Federal Institute of}\\{\em Technology (EPFL),
CH-1015 Lausanne, Switzerland}\\{\em and}\\
{\em Institute for Nuclear Research of the Russian Academy of Sciences,}\\
{\em 60th October Anniversary prospect 7a, Moscow 117312, Russia}}
\date{January 16, 2004}
\begin{document}
\maketitle

\begin{abstract}
We consider four-dimensional massive gravity with the Fierz--Pauli
mass term. The analysis of the scalar sector has revealed recently that
this theory becomes strongly coupled above the energy scale
$\Lambda = (M_{Pl}m^4)^\frac{1}{5}$ where $m$ is the mass of the graviton.
We confirm this scale by explicit calculations of the four-graviton
scattering amplitude and of the loop correction to the interaction between
conserved sources.
\end{abstract}

\section{Introduction and summary}

In view of the evidence for the accelerated expansion of the Universe,
it is of interest to understand whether there exist consistent and
phenomenologically acceptable gravitational theories which deviate from
general relativity at cosmological scales. The simplest possibility would be
to give a mass to the graviton by adding the Fierz--Pauli term~\cite{pauli}
to the Einstein action. At the linearized level, this would modify classical
gravity even at energy scales exceeding the graviton mass, due to the
van Dam--Veltman--Zakharov phenomenon~\cite{vandam}, but this undesirable
feature may be cured by non-linear effects~\cite{vainshtein, deffayet}.
More serious problem emerges at quantum level: it has been
argued~\cite{arkani} that the theory becomes strongly coupled above the energy
scale \begin{equation}\label{scale}
\Lambda = (M_{Pl}m^4)^{\frac{1}{5}} \end{equation}
where $m$ is the graviton mass. For $m \sim H_0$, the present value of the
Hubble parameter, this scale is unacceptably low. Possible higher order terms
do not improve the ultraviolet behaviour to phenomenologically acceptable
level~\cite{arkani}.
Similar problem is inherent also in brane-world
models~\cite{charmousis, kogan, DGP} with gravity modifed at ultra-large
distances: they either are strongly coupled at unacceptably low
energies~\cite{luty, VR, dubovsky, chacko} or have
ghosts~\cite{pilo, luty, dubovsky, chacko}.
A possible way out is related to the breaking of the
Lorentz invariance, as suggested recently~\cite{newarkani}.

\begin{figure}[h]
\begin{center}
\includegraphics[width=3cm]{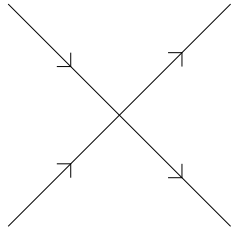}
\caption{Scattering through the four point vertex.}
\label{scat4}
\end{center}
\end{figure}

\begin{figure}[h]
\begin{center}
\includegraphics[width=12cm]{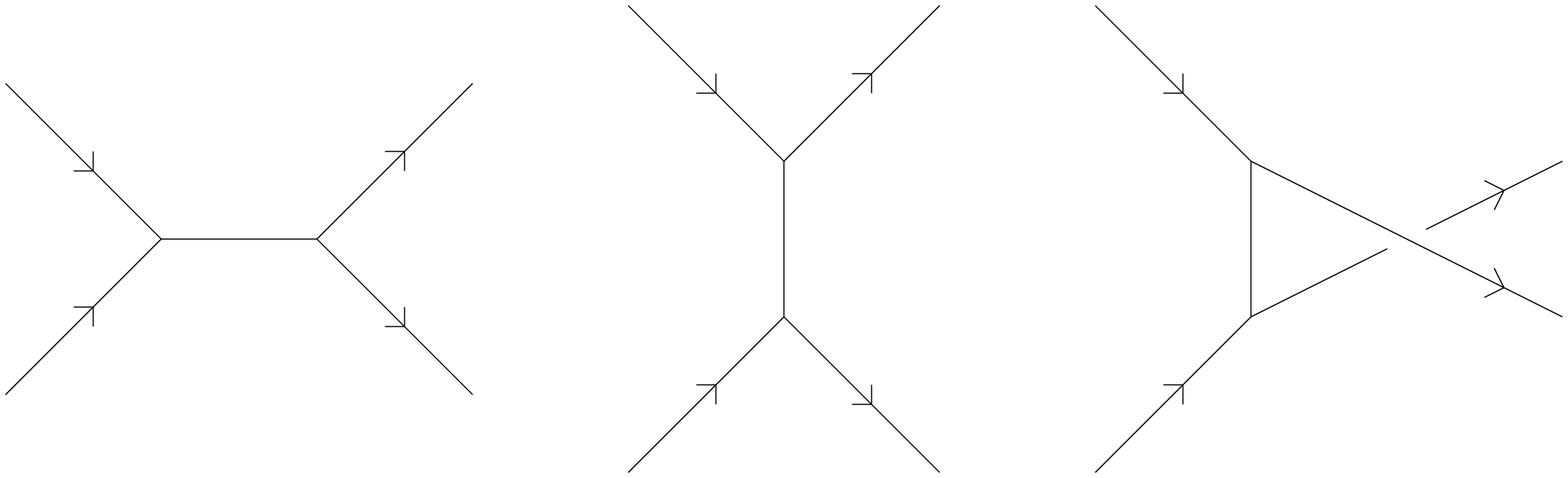}
\caption{Scattering via the s-, t- and u-channel exchange.}
\label{scat3}
\end{center}
\end{figure}

\begin{figure}[ht]
\begin{center}
\input{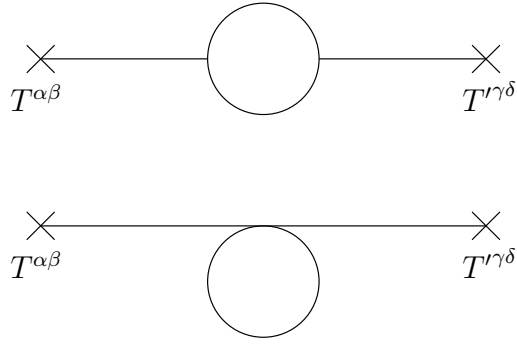}
\caption{Graviton propagator at one loop.}
\label{interact}
\end{center}
\end{figure}

The energy scale~(\ref{scale}) was derived in Ref.~\cite{arkani} in a
somewhat indirect way, namely, by making use of a gravitational analogue of the
sigma-model approach. It is of interest to see this scale directly in
scattering amplitudes and loop corrections. This is precisely the purpose of
this paper: we calculate directly the $2\rightarrow 2$ scattering amplitude
of longitudinal gravitons (Figs.~\ref{scat4} and \ref{scat3})
and the loop correction to the graviton propagator
(Fig.~\ref{interact}), both in flat background.
We confirm that the strong coupling scale in massive gravity is given by
Eq.~(\ref{scale}). This scale appears in the two calculations in a way which
is not completely trivial: a na\"{\i}ve power couting would suggest even
lower energy scale, but two leading orders in $m^{-1}$ cancel out in both
cases. As an example, the wave function of the longitudinal graviton is
proportional to $m^{-2}$, while the largest term in the graviton propagator
is proportional to $m^{-4}$. Thus, na\"{\i}ve power couting would suggest
that the diagrams of Fig.~\ref{scat3} are of the order
$\frac{E^{14}}{M^2_{Pl}m^{12}}$. However, on-shell amplitude is in fact of
order $\frac{E^{10}}{M^2_{Pl}m^{8}}$, due to cancellations. This
immediately implies that the strong interaction scale is indeed given by
Eq.~(\ref{scale}).

\section{Massive graviton wave functions and propagator}

The graviton field $h_{\mu\nu}(x)$ is the perturbation about
the Minkowski metric $\eta_{\mu\nu} = diag(1,-1,-1,-1)$,
\begin{equation}
h_{\mu\nu}(x) = g_{\mu\nu}(x) - \eta_{\mu\nu}
\end{equation}
The graviton is given a mass by adding the Fierz-Pauli term~\cite{pauli}
to the Einstein action,
\begin{equation} \label{action}
S = - \frac{M_{Pl}^2}{16 \pi} \int d^4x \sqrt{-g} \ {\cal R} \ \ 
- f^4 \!\int d^4x \left( h_{\mu\nu} h^{\mu\nu}
- h^\mu_{\ \mu} h^\nu_{\ \nu} \right)
\end{equation}
where the indices of $h_{\mu\nu}(x)$ are raised by the Minkowski metric.

By solving the linearised field equation, one finds that the graviton field
has a mass, $m^2=\frac{64\pi f^4}{M_{Pl}^2}$, and is decomposed as follows,
\begin{equation}
h_{\mu\nu}(x) = \frac{4 \sqrt{2\pi}}{M_{Pl}} \int
\frac{d^3k}{(2\pi)^3(2k^0)} \sum_{\alpha = 1}^{5}
\left( e_{\mu\nu}^{(\alpha)} a^\alpha ({\bf k}) e^{-ikx} + h.c. \right)
\end{equation}
where the five polarisation tensors have
the following proprieties,
\begin{displaymath}
e_{\mu\nu}^{(\alpha)}=e_{\nu\mu}^{(\alpha)} \quad ; \quad
e_{\quad \ \mu}^{(\alpha) \mu}=0 \quad ; \quad
k_\mu e^{(\alpha)\mu\nu}=0 \quad ; \quad
e_{\mu\nu}^{(\alpha)} e^{(\beta)\mu\nu}=\delta^{\alpha\beta}
\end{displaymath}
and the creation and annihilation operators are normalized in the standard way,
\begin{displaymath}
[a^\alpha({\bf k}), a^{\beta\dagger} ({\bf k'}) ] = (2\pi)^3 (2k^0)
\delta^3({\bf k}-{\bf k'}) \delta^{\alpha\beta}
\end{displaymath}

The five polarisation tensors $e_{\mu\nu}^{(\alpha)}$ of the massive
graviton can be expressed in terms of the three polarisation
vectors $e_\mu^{(i)}$ of the massive vector field,
\begin{eqnarray}
e_{\mu\nu}^{(1)}=\frac{1}{\sqrt{2}}
(e_\mu^{(1)}e_\nu^{(2)} + e_\mu^{(2)}e_\nu^{(1)}) \quad&;&\quad
e_{\mu\nu}^{(2)}=\frac{1}{\sqrt{2}}
(e_\mu^{(1)}e_\nu^{(1)} - e_\mu^{(2)}e_\nu^{(2)}) \nonumber\\
e_{\mu\nu}^{(3)}=\frac{1}{\sqrt{2}}
(e_\mu^{(1)}e_\nu^{(3)} + e_\mu^{(3)}e_\nu^{(1)}) \quad&;&\quad
e_{\mu\nu}^{(4)}=\frac{1}{\sqrt{2}}
(e_\mu^{(2)}e_\nu^{(3)} + e_\mu^{(3)}e_\nu^{(2)}) \nonumber\\
e_{\mu\nu}^{(5)}=\frac{1}{\sqrt{6}} (e_\mu^{(1)}e_\nu^{(1)}\!\!\!\!
&+&\!\!\!\! e_\mu^{(2)}e_\nu^{(2)} -2 e_\mu^{(3)}e_\nu^{(3)})
\label{polaris}\end{eqnarray}
where the two first polarisation tensors (as well as the two first
polarisation vectors) are the ones of the massless case.

In a frame where $k^\mu = \left(k^0, 0, 0, k^3\right)$, one has
\begin{eqnarray*}
e_\mu^{(1)} = \left(0, 1, 0, 0\right) \quad ; \quad
e_\mu^{(2)} = \left(0, 0, 1, 0\right) \quad ; \quad
e_\mu^{(3)} = \frac{1}{m}\left(k^3, 0, 0, -k^0\right)
\end{eqnarray*}
Hence, the longitudinal
tensor $e_{\mu\nu}^{(5)}$ is the most singular one as $m\rightarrow 0$.
For further calculations it is convenient to express it in terms of momentum,
\begin{equation} \label{polapprox}
e_{\mu\nu}^{(5)} = -\sqrt{\frac{2}{3}}\ \frac{k_\mu k_\nu}{m^2}
+ {\cal O} (m^0)
\end{equation}

The graviton propagator obeys the following equation,
\begin{eqnarray*}
\frac{M_{Pl}^2}{64 \pi}\Big[ (
\eta_{\gamma\alpha}\eta_{\delta\beta}+\eta_{\gamma\beta}\eta_{\delta\alpha}
-2\eta_{\gamma\delta}\eta_{\alpha\beta})(k^2-m^2)&&\\
-(k_\delta\eta_{\gamma\alpha}k_\beta+k_\delta\eta_{\gamma\beta}k_\alpha
+k_\gamma\eta_{\delta\alpha}k_\beta+k_\gamma\eta_{\delta\beta}k_\alpha)&&\\
+2(\eta_{\gamma\delta}k_\alpha k_\beta+k_\gamma k_\delta\eta_{\alpha\beta})
\Big]G^{\alpha\beta\mu\nu}(k)&=&\frac{i}{2}
\left( \delta_\gamma^\mu \delta_\delta^\nu
+ \delta_\gamma^\nu \delta_\delta^\mu \right)
\end{eqnarray*}
One finds that the massive graviton propagator contains terms of order
$1$, $m^{-2}$ and $m^{-4}$,
\begin{eqnarray} \label{prop}
G_{\alpha\beta\mu\nu}(k)&=&\frac{16 \pi i}{M_{Pl}^2 (k^2 - m^2)}
[ \eta_{\alpha\mu}\eta_{\beta\nu} + \eta_{\alpha\nu}\eta_{\beta\mu}
-\frac{2}{3}\eta_{\alpha\beta}\eta_{\mu\nu} \nonumber\\
& &- \frac{1}{m^2} (k_\beta\eta_{\alpha\mu}k_\nu+
k_\beta\eta_{\alpha\nu}k_\mu+k_\alpha\eta_{\beta\mu}k_\nu+
k_\alpha\eta_{\beta\nu}k_\mu) \nonumber\\
& &+\frac{2}{3m^2} (\eta_{\alpha\beta}k_\mu k_\nu
+ k_\alpha k_\beta\eta_{\mu\nu})
+ \frac{4}{3m^4} k_\alpha k_\beta k_\mu k_\nu ]
\end{eqnarray}
The propagator can also be written as follows,
\begin{eqnarray}
G_{\alpha\beta\mu\nu}(k)&=&\frac{16 \pi i}{M_{Pl}^2 (k^2 - m^2)}
[ \hat{\eta}_{\alpha\mu}\hat{\eta}_{\beta\nu}
+ \hat{\eta}_{\alpha\nu}\hat{\eta}_{\beta\mu}
-\frac{2}{3}\hat{\eta}_{\alpha\beta}\hat{\eta}_{\mu\nu} ]
\end{eqnarray}
where $\hat{\eta}_{\mu\nu}=\eta_{\mu\nu}-\frac{k_\mu k_\nu}{m^2}$.
The coefficient $-\frac{2}{3}$ here is different from $-1$ of the massless
case; this is precisely the van Dam--Veltman--Zakharov
discontinuity~\cite{vandam}.

As a cross check, near mass shell one has
\begin{displaymath}
G_{\alpha\beta\mu\nu}(k)=\frac{32 \pi i}{M^2_{Pl}(k^2-m^2)}
\sum_{\gamma = 1}^{5}e_{\alpha\beta}^{(\gamma)} e_{\mu\nu}^{(\gamma)}
\end{displaymath}
which is precisely the expected contribution of the nearly on-shell gravitons.

\section{Scattering}

Interaction between gravitons comes from non-linear terms in the Einstein
action. The corresponding three-point and four-point vertices are given in
Appendix. We use them for calculating the amplitude of scattering of two
massive gravitons; at the tree level it is given by the sum of the diagrams
shown in Figs.~\ref{scat4} and \ref{scat3}.

By na\"{\i}ve power counting, one might think that the term of order
$m^{-4}$ in the massive propagator~(\ref{prop}) dominates
the scattering amplitude.
However, once the explicit form of the vertex~(\ref{3pt}) is used, and
external legs are taken on-shell, its contribution in fact picks up a factor
$m^4$. Likewise, the term in the propagator with $m^{-2}$ in front picks up a
factor $m^2$, so effectively the propagator is of order $m^0$.
These cancellations occur for each diagram in Fig.~\ref{scat3} separately,
and for all polarisations of gravitons in the legs, provided these are
on-shell. Then, the inverse powers of the mass $m$ in the scattering
amplitude come only from the polarisation tensor $e_{\mu\nu}^{(\alpha)}$.

Therefore, the largest amplitude at $E\gg m$ involves longitudinal gravitons
in all four legs. Making use of Eq.~(\ref{polapprox}), we obtain for the
diagram of Fig.~\ref{scat4}, the diagram of Fig.~\ref{scat3} and their sum:
\begin{eqnarray} \label{ampl4}
{\cal M}_{4} &=& -\frac{2\pi}{9 M_{Pl}^2 m^8} stu (s^2+t^2+u^2)\\
\label{ampl3}
{\cal M}_{3} &=& \frac{7\pi}{54 M_{Pl}^2 m^8} stu (s^2+t^2+u^2)\\
\label{ampltot}
{\cal M}_{TOT} &=& -\frac{5\pi}{54 M_{Pl}^2 m^8} stu (s^2+t^2+u^2)
\end{eqnarray}
where $s$, $t$ and $u$ are the Mandelstam variables.
These expressions are symmetric in $s$, $t$ and $u$ as they should.

Thus, at high energies the $2\rightarrow 2$ amplitude for longitudinal
gravitons is of the order
\begin{displaymath}
{\cal M} \ \sim \ \frac{E^{10}}{\Lambda^{10}}
\end{displaymath}
where $\Lambda$ is precisely the scale~(\ref{scale}). The standard unitarity
argument implies then that the theory is strongly coupled at
$E\gsim \Lambda$, in agreement with Ref.~\cite{arkani}.

\section{One-loop graviton propagator}

Another way to see the strong coupling scale is to study the loop correction
to the graviton propagator. We do this by adding external sources of the
form $T^{\alpha\beta}h_{\alpha\beta}$ and
${T'}^{\gamma\delta}h_{\gamma\delta}$, and calculating the interaction
between $T^{\alpha\beta}$ and ${T'}^{\gamma\delta}$.
At the tree level, the interaction between two symmetric conserved sources
$T^{\alpha\beta}(x)$ and ${T'}^{\gamma\delta}(x)$ is
\begin{eqnarray} \label{newtonmassive}
\!\!\!\!\!\!\!\int &&\!\!\!\!\!\!\!\!\!\!\!\!\!\! \frac{d^4k}{(2\pi)^4}
\tilde{T}^{\alpha\beta}(k)
G_{\alpha\beta\gamma\delta}^{(0)}(k)
{{\tilde{T}}'}{}^{\gamma\delta}(-k)\nonumber\\&=&\!\!
\int \frac{d^4k}{(2\pi)^4} 
\frac{32 \pi i}{M_{Pl}^2 (k^2 - m^2)}
\left(\tilde{T}^{\alpha\beta}(k){{\tilde{T}}_{\alpha\beta}'}(-k)-\frac{1}{3}
\tilde{T}^\alpha_{\ \alpha}(k){{\tilde{T}}'}{}^\beta_{\,\beta}(-k)\right)
\end{eqnarray}

We now calculate the interaction at the one loop order; the corresponding
diagrams are shown in Fig.~\ref{interact}.
We are interested in the regime $k^2 \gg m^2$.

As the sources are transverse, all terms in the propagator
enhanced by $m^{-4}$ and $m^{-2}$ vanish upon contracting with
$T^{\alpha\beta}$ or ${T'}^{\alpha\beta}$, exept for one term.
This non-vanishing term is of the form $\eta_{\alpha\beta}k_\mu k_\nu$,
where the two indices of the Minkowski metric are contracted with the
energy-momentum tensor.
So, one might think that for the first diagram of Fig.~\ref{interact}, the two
propagators in the loop each contribute as $m^{-4}$, and the
two external propagators each contribute as $m^{-2}$.
This would lead to a result of the order $m^{-12}$,
but as in the previous section there are cancellations.
Unlike in the previous section, however, these cancellations have nothing to
do with the mass-shell condition. By direct calculation one finds that the
terms of order $m^{-12}$ and $m^{-10}$ vanish, so one is left with
contribution of order $m^{-8}$ from the first diagram of Fig.~\ref{interact}.
The second diagram has only one propagator inside the loop, so it is at most
of order $m^{-8}$.

To check that the terms of order $m^{-8}$ do not cancel out, it is sufficient
to consider traceless sources,
\begin{displaymath}
T^\alpha_{\ \alpha}={T'}^\alpha_{\ \alpha}=0
\end{displaymath}
This simplifies algebra considerably (in particular, the second diagram in
Fig.~\ref{interact} does not contribute to the order $m^{-8}$), and we
obtain for the sum of the tree level and one-loop interactions,
\begin{eqnarray} \label{masslessinteract}
\int &&\!\!\!\!\!\!\!\!\!\!\!\!\! \frac{d^4k}{(2\pi)^4}
\tilde{T}^{\alpha\beta}(k)
\Big[G_{\alpha\beta\gamma\delta}^{(0)}(k)+
G_{\alpha\beta\gamma\delta}^{(1)}(k)\Big]
{{\tilde{T}}'}{}^{\gamma\delta}(-k)\nonumber\\&=&
\int \frac{d^4k}{(2\pi)^4} \tilde{T}^{\alpha\beta}(k)
{{\tilde{T}}_{\alpha\beta}'}(-k)
\frac{32 \pi i}{M_{Pl}^2 (k^2 - m^2)}\nonumber\\&&
\left(1+\frac{k^{10}}{2160\pi M_{Pl}^2 m^8} \log(k^2) +
\frac{P(k)}{M_{Pl}^2 m^8}+\frac{{\cal O} (m^{-6})}{M_{Pl}^2}\right)
\end{eqnarray}
where $P(k)$ is a polynomial in $k$.

Thus, the correction of order $m^{-8}$ does not cancel out
in the graviton propagator, even for traceless sources.
This correction becomes comparable to the tree level term at the
energy scale $\Lambda$, Eq.~(\ref{scale}). This again demonstrates that
$\Lambda$ is indeed the strong coupling scale in massive gravity.

\paragraph{Acknowledgment:}

The author would like to thank V.~Rubakov and S. Dubovsky for useful
discussions.

\appendix
\section{Vertices}

To obtain the expression for the  Einstein action (\ref{action}) at the third
and fourth order in $h_{\mu\nu}$, one makes use the
following expansions,
\begin{eqnarray}
\sqrt{-g} &=& 1 + \frac{1}{2} h^\alpha_{\ \alpha}
+ \frac{1}{8} (h^\alpha_{\ \alpha})^2
- \frac{1}{4} h_{\alpha\beta} h^{\alpha\beta}
+ \frac{1}{48} (h^\alpha_{\ \alpha})^3 \nonumber\\
& &- \frac{1}{8} h^\alpha_{\ \alpha} h_{\beta\gamma} h^{\beta\gamma}
+ \frac{1}{6} h_{\alpha\beta} h^{\beta\gamma} h_\gamma^{\ \alpha}
+ {\cal O} (h_{\alpha\beta}^4)\\
g^{\mu\nu} &=& \eta^{\mu\nu} - h^{\mu\nu} + h^\mu_{\ \sigma}h^{\sigma\nu}
- h^{\mu\sigma} h_{\sigma\tau} h^{\tau\nu} + {\cal O} (h_{\mu\nu}^4)
\end{eqnarray}
With these, one finds the expressions for the three and
four point vertices (cf. Ref.~\cite{dewitt}):
\begin{eqnarray}
V^{(\mu\nu)(\sigma\tau)(\rho\lambda)}_3 &=& -\frac{iM_{Pl}^2}{16\pi}
\textrm{Sym} \Bigg\{-\frac{1}{4}P_3\left[\eta^{\mu\nu}\eta^{\sigma\tau}
\eta^{\rho\lambda}k_1\cdot k_2\right]\nonumber\\&&
+\frac{1}{4}P_3\left[\eta^{\mu\nu}\eta^{\sigma\rho}
\eta^{\tau\lambda}(k_2\cdot k_3-2k_1\cdot k_1)\right]\nonumber\\&&
-P_3\left[\eta^{\lambda\mu}\eta^{\nu\sigma}
\eta^{\tau\rho}k_1\cdot k_2\right]
-\frac{1}{4}P_6\left[\eta^{\mu\nu}\eta^{\sigma\tau}
k_1^\rho k_1^\lambda\right]\nonumber\\&&
+P_3\left[\eta^{\mu\nu}\eta^{\sigma\rho}
(k_1^\tau k_1^\lambda -\frac{1}{2} k_3^\tau k_2^\lambda)\right]
+\frac{1}{2}P_6\left[\eta^{\mu\sigma}\eta^{\nu\tau}
k_1^\rho k_1^\lambda\right]\nonumber\\&&
+\frac{1}{2}P_3\left[\eta^{\mu\sigma}\eta^{\nu\tau}
k_1^\rho k_2^\lambda\right]
+P_3\left[\eta^{\mu\sigma}\eta^{\nu\rho}k_3^\tau k_2^\lambda\right]
\nonumber\\&&
+P_6\left[\eta^{\mu\sigma}\eta^{\nu\rho}k_1^\tau k_2^\lambda\right]
\Bigg\}\label{3pt}
\end{eqnarray}
\begin{eqnarray}
V^{*(\mu\nu)(\sigma\tau)(\rho\lambda)(\pi\xi)}_4 &=& -\frac{iM_{Pl}^2}{16\pi}
\textrm{Sym}
\Bigg\{\frac{1}{4}P_3\left[\eta^{\mu\sigma}\eta^{\nu\tau}\eta^{\rho\pi}
\eta^{\lambda\xi}(k_1\cdot k_2+2m^2)\right]\nonumber\\
&&-2P_3\left[\eta^{\xi\mu}\eta^{\nu\sigma}\eta^{\tau\rho}
\eta^{\lambda\pi}m^2\right]
-\frac{1}{2}P_{12}\left[\eta^{\xi\sigma}\eta^{\tau\rho}\eta^{\lambda\pi}
k_2^\mu k_2^\nu\right]\nonumber\\
&&-\frac{1}{2}P_{12}\left[\eta^{\mu\sigma}\eta^{\nu\tau}\eta^{\rho\pi}
k_1^\lambda k_1^\xi\right]\nonumber\\
&&+P_{12}\left[\eta^{\sigma\rho}\eta^{\lambda\pi}\eta^{\xi\mu}
(k_2^\nu k_1^\tau -k_3^\nu k_4^\tau)\right]
\Bigg\}\label{4pt}
\end{eqnarray}
where $k_1+k_2+k_3=0$ for the three point vertex and $k_1+k_2+k_3+k_4=0$
for the four point vertex.
The symbol $P$ means that one has to sum over all distinct
permutations of the triplets of indices $1\mu\nu$, $2\sigma\tau$,
$3\rho\lambda$ and $4\pi\xi$.
The subscript of $P$ indicates the number of distinct permutations over which
the summation has to be carried out.
The ``Sym'' symbol means that the total expression has to be symmetrized
in each pair of indices $(\mu\nu)$, $(\sigma\tau)$, $(\rho\lambda)$
and $(\pi\xi)$.
In fact, this symmetrization is not required as long as one only multiplies
the vertices by the propagator (\ref{prop}) or the polarisation
tensors (\ref{polaris}) which are already symmetric.

The ${}*$ symbol indicates that, for the four point vertex (\ref{4pt}),
the four legs are on mass shell.
The general off-shell expression of the four point vertex is given in
Ref.~\cite{dewitt}.

\end{document}